\documentclass[twocolumn,amssymb,aps,pre,showpacs,nofootinbib,10pt, superscriptaddress]{revtex4-1}
\usepackage{graphicx}
\usepackage{amsmath}
\usepackage[caption=false,font=normalsize]{subfig}
\usepackage{float}
\usepackage{nomencl}
\usepackage{bigints}

\usepackage{hyperref}
\usepackage{dcolumn}
\usepackage{bm}
\usepackage{latexsym}
\usepackage{amssymb}

\usepackage{commath}
\usepackage[caption=false]{subfig}
\usepackage{color}

\newcommand{\valuegiv}{0.2675}

\usepackage{titlesec}
\titlespacing*{\section}{0pt}{1.1\baselineskip}{\baselineskip}
\titlespacing*{\subsection}{0pt}{1.1\baselineskip}{\baselineskip}
\usepackage{indentfirst}

\newlength{\nomitemorigsep}
\usepackage{makeidx}

\begin{document}
\title{Analysis of the Dynamics and Topology Dependencies of Small Perturbations in Electric Transmission Grids}
\author{Luiscarlos~A.~Torres-S\'anchez}
\email{l.torres-sanchez@outlook.com}
\affiliation{%
    Department of Energy Division, decon international GmbH, 61352 Bad Homburg, Germany.
}

\author{Giuseppe T. Freitas de Abreu}
\email{g.abreu@jacobs-university.de; g-abreu@fc.ritsumei.ac.jp}
\affiliation{%
    Department of Computer Science and Electrical Engineering, Jacobs University
Bremen, 28759 Bremen, Germany.
}

\affiliation{%
    Department of Electrical
and Electronic Engineering, Ritsumeikan University, Kusatsu 525-8577, Japan.
}

\author{Stefan Kettemann}
\email{s.kettemann@jacobs-university.de}
\affiliation{%
     Department of Physics and Earth Science, Jacobs University Bremen, 28759 Bremen, Germany.
}
\affiliation{
    Division of Advanced
Materials Science, Pohang University of Science and Technology, Pohang 790-784, South Korea.
}

\date{\today}

\begin{abstract}
 
Through an eigenanalysis of small perturbations, as typically done in small-signal stability studies, we intend to discover the underlying reasons that make those perturbations propagate in some way or another in the grid. To this end, we establish connections between the perturbations time-scale and topological metrics. Namely, the algebraic connectivity and the Fiedler vector of a generalized/weighted Laplacian matrix that depends on the stationary phase solutions of the system and is thereby inherently conditioned by the topology and the power distribution. Then, we aim to find out the isolated influence of topology on the perturbations when the network interacting agents have, in principle, opposite behaviors (i.e. producers and consumers). To do so, we study three networks: Small-world, Random, German grid. Furthermore, we tackle the effect of machine clustering on small perturbations and the influence of the network's average clustering coefficient on the intensity localization of the generalized Fiedler vector. Finally, we propose ways in which future (dynamic topology control) and existing (power system stabilizer) grid control strategies can adapt their response to comprehensively consider the topology and remote signals in the system. 
\end{abstract}


\maketitle



%


\makenomenclature
\renewcommand{\nomgroup}[1]{%
  \itemsep\nomitemorigsep%
  \ifthenelse{%
    \equal{#1}{A}%
  }{%

 }{%
    \ifthenelse{\equal{#1}{B}}{%

    }{}%
  }%
  \itemsep\nomitemsep
} 
\nomenclature{$\varphi_i$}{Rotor angle at node $i$ in $rad$}
\nomenclature{$J$}{Moment of inertia in $kgm^2$}
\nomenclature{$\gamma$}{Damping coefficient in $Nms$}
\nomenclature{$K_{ij}$}{Power line capacity between nodes $i$ and $j$ in $W$}
\nomenclature{$\omega$}{Grid angular frequency in $rad/s$}
\nomenclature{$\theta_i$}{Phase shift at node $i$ in $rad$}
\nomenclature{$\alpha_i$}{Rotor angle perturbation at node $i$ in $rad$}
\nomenclature{$A$}{Unweighted Graph Adjacency matrix}
\nomenclature{$L$}{Unormalized Unweigthed Graph Laplacian matrix}
\nomenclature{$B$}{Oriented Incidence matrix}
\nomenclature{$d_i$}{Degree of node $i$}
\nomenclature{$C_i$}{Clustering coefficient of node $i$}
\nomenclature{$E$}{Coupling matrix, Generalized Laplacian matrix}
\nomenclature{$t_{ij}$}{Coupling amplitude between nodes $i$ and $j$}
\nomenclature{$\Gamma$}{Perturbation's relaxation/damping rate in $Hz$}
\nomenclature{$\l_{ij}$}{Shortest distance between nodes $i$ and $j$}
\nomenclature{$D$}{Node-degree matrix}
\nomenclature{$\epsilon_k$}{Real eigenfrequency $k$ in $Hz$}
\nomenclature{$c_{ik}$}{Element $i$ of eigenmode/eigenvector $k$}
\nomenclature{$\Omega_k$}{Complex eigenfrequency $k$ in $Hz$}
\nomenclature{$\kappa$}{Number of edges in the network}
\nomenclature{$N$}{Number of nodes in the network}
\nomenclature{$p$}{Rewiring probability for the Watts-Strogatz algorithm}
\nomenclature{$h_i$}{Number of links shared among the $d_i$ neighbors of $i$}
\nomenclature{$P_i$}{Mechanical power at node $i$ in $W$}
\nomenclature{$P_{e_i}$}{Electric power at node $i$ in $W$}
\nomenclature{$a(G)$}{Algebraic connectivity of graph $G$}
\nomenclature{$b_{k \sigma_k}$}{Fourier series expansion coefficients}
\printnomenclature

\section{Introduction}


 \indent Electric
power grids   provide a highly reliable electrical service
  to billions of customers.
   In fact,  the average outage time experienced by a consumer 
 has kept  decreasing in recent  years, reaching a record low of $12.5$ minutes in Germany, in $2014$ \cite{bna}. However, the energy transition from a centralized power production with unilateral power flow towards an increased supply of decentralized and more volatile  renewable energy resources with bidirectional flow, might become harmful for the stability of electricity grids in the future.
   In  the currently existing grids, the synchronous 
    generators  and synchronous motors
    provide, with their rotating masses, high inertia to the system, which automatically 
     reacts to disturbances \cite{meier,kundur_ieee}.     
     For instance,  an abrupt  increase in  load demand 
    can be momentarily balanced by a change of the kinetic energy of  
     rotating synchronous generators, 
       causing some generators to slow down and deviate from 
        the grid frequency,  but ensuring the overall stability of the network.
       With an increasing share of renewable energy,  this  buffer 
        for the electrical energy  is expected to decrease 
         since solar cells and conventional wind turbines  do not  provide such  inertia to the system \cite{ulbig}. 
          Therefore, it will be increasingly important to  obtain a deeper 
           understanding of
            how fast disturbances decay and  spread  in the grid and how this 
              depends on the topological connectivity and the system parameters, 
              in order 
               to maintain a reliable control of the network.

Many authors have studied the role of system topology for the robustness of power grids against large disturbances, such as intentional and random removals of nodes and edges \cite{cuadra, timme,jung}. For small disturbances, on the other hand, the small-disturbance rotor angle stability has been properly defined \cite{kundur_ieee} and thoroughly studied by assessing the solution of the system swing equations and its conditions of stability. In fact, extensive attention has been given to the eigenvalues and eigenvectors of the 
 stability matrix of multiple-machine systems \cite{timme,machowski, coletta} to, for instance, optimize the parameters and grid location of Power System Stabilizers (PSS) \cite{machowski}. Nonetheless, little attention has been given to the 
  propagation  of small disturbances and 
   how the latter depends on   grid topology and the distribution of  system parameters. 
   In order to study the decay and propagation 
    of disturbances, we  implement a hybrid approach  to combine graph theory tools with electric parameters of inductive grids \cite{cuadra} and consider only undirected graphs to depict the smart grid concept, according to which consumers could rapidly become producers and exchange the existing hierarchical power transmission into a bidirectional system.
    
    In this article, we first introduce the mathematical model for generators, loads and perturbations. Secondly, we explain the construction of three networks (Small-world, Random, Geman grid) and how their properties relate to the perturbations dynamics. Then, we perform a spectral analysis of the perturbations eigenfrequency distributions and eigenvectors localization in these topologies. Finally, in an attempt to highlight the crucial connection to topology, we analyze the responses of the dynamic topology control strategy and the power system stabilizers.
\section{Phase Dynamics Analysis}

\subsection{Mathematical Model}

Phase  dynamics in AC electricity grids have been modeled by active power balance equations with  additional  terms  describing the dynamics of rotating machines \cite{timme, hill,filatrella,schmietendorf,heuck,menck}. We specifically assume loads to be synchronous motors whose $\varphi_i$ dynamics can be modeled by the swing equation of synchronous generators \cite{motter}. This second-order differential equation describes the inertia to changes in kinetic energy through $J$ and $\gamma$. Adding these terms to the active power balance equations yields, for purely inductive transmission lines \cite{timme, hill,filatrella,heuck,menck},
    \begin{equation} \label{dynamic0}
        P_i =\left( \frac{J}{2}  \frac{d}{dt}  + \gamma \right)  \left(\frac{d \varphi_i}{d t} \right)^2  + \underbrace{\sum_j K_{ij} \sin{(\varphi_i-\varphi_j)}}_{P_{e_i}}.
    \end{equation}

Eq.(\ref{dynamic0}) is analog to an unregulated generator, where primary and secondary 
 frequency control and  voltage regulator actions are disregarded or considered to have large time constants. Therefore, we can assume $P_i$ and the excitation voltages to be constant in time \cite{machowski}. Eq.(\ref{dynamic0}), with $K_{ij}=K A_{ij}$,  corresponds to a homogeneous system in which all generators, motors, and transmission lines, have identical inertia, damping and power line capacity parameters. 
 
 A homogeneous system allows to focus on the influence of network topology \cite{motter} and can also be used to model a grid with low level of inertia (produced by the integration of renewable energy sources), whose frequency deviations are controlled by a simplistic consideration of fast primary control (e.g. Battery Energy Storage Systems). This control response can  be simply modeled as an additional damping term \cite{ulbig}. Here, we consider fixed voltages (i.e. at $V_i= 1$ p.u.), which eliminates dynamic terms in the reactive power balance equation
     as they only appear in higher order
      when voltage dynamics- in addition to phase dynamics- 
        are considered \cite{machowski,schmietendorf, kundur_book}.
        
The rotor angle is expressed as $\varphi_i(t) =  \omega   t + \theta_i(t)$. By assuming that $\dot{\theta_i}<<\omega$ and that the rate at which energy is stored in the kinetic term is much less than the rate at which energy is dissipated by friction (i.e.  $|J\ddot{\theta}_i|<<2\gamma\omega$), Eq.(\ref{dynamic0}) can be simplified as \cite{filatrella}: 
       
        \begin{equation} \label{dynamic}
        P_i =J \omega \ddot{\theta_i}+2\gamma\omega\dot{\theta_i}+ K\sum_j A_{ij} \sin{(\theta_i-\theta_j)}.
       \end{equation}

      \subsection{Dynamics of Disturbances in the Grid:}
      
 In order to study the propagation of disturbances, we set  $ \varphi_i(t) =  \omega   t + \theta^0_i +  \alpha_i(t)$ 
 with steady state phases  $\theta^0_i$, the
solutions of Eq.(\ref{dynamic}). 
   The dynamics of $\alpha_i(t)$ are governed by: 
        \begin{equation} \label{alphaharmonic3}
        \begin{split}
             \partial_t^2   \alpha_i  + 
                  2 \Gamma  \partial_t   \alpha_i   =& \frac{P_{i}}{J \omega}\\ &- \sum_j \frac{K}{J \omega} A_{ij} \sin   (\theta^0_i
       - \theta^0_j + \alpha_i -\alpha_j),
       \end{split}
          \end{equation}       
          where $\Gamma = \gamma/J$. Since the steady-state natural or inherent stability of a system can be analyzed via a linearized unregulated condition \cite{machowski}, we consider  small  perturbations from  the  stationary state, as typically done in small-signal stability analyses, and expand  Eq.(\ref{alphaharmonic3}) in ($\alpha_i -\alpha_j$),
        which yields  linear equations on the grid \cite{kettemann}:
    \begin{equation} \label{alphaharmonic4}
    \partial_t^2   \alpha_i  + 2 \Gamma  \partial_t   \alpha_i 
                   = -
       \sum_j   t_{ij}    (\alpha_i -\alpha_j),
          \end{equation}
         with $t_{ij} = \frac{K}{J\omega} A_{ij}\cos(\theta^0_i - \theta^0_j)$ \cite {kettemann}. We let $E$ to be formed as $E_{ii}=\sum_{j} t_{ij}$, and $E_{ij}=-t_{ij}$. This matrix 
         is a weighted Laplacian and 
         has been previously identified in synchronization studies of coupled-oscillator networks, with possibly including ohmic losses \cite{myers}, as well as in linear stability studies for purely inductive grids \cite{coletta}, under the name of  \textit{stability matrix}. 
         
          We express the perturbation at node $i$ as a complex Fourier series, 
           $\alpha_i (t)   =  \sum_{k=1, \sigma_k = \pm}^N b_{k \sigma_k}
               c_{i k} \exp {(-j \Omega_{k \sigma_k} t)}$,
           to then plug it into
               Eq.(\ref{alphaharmonic4}) to obtain:
                 \begin{eqnarray} \label{randommatrix1}
                    (\Omega_{k \sigma_k}^2 +  j2\Gamma \Omega_{k \sigma_k})c_{ik}  
                   =      
       E \vec c_k.
          \end{eqnarray}

      The stationary solution of the perturbation is 
$\vec{c_1}=(1/\sqrt{N}) {\textbf 1}$ and $\epsilon_1=0$. This holds for both unweighted and weighted Laplacian matrices due to the linear dependence of the diagonal on the off-diagonal elements.       
For $\Gamma=0$, we find from the eigenvalue equation $E\vec{c}_k= \Lambda_k\vec{c}_k$, the eigenvectors $\vec{c}_k$ and the eigenvalues $\Lambda_{k}$ of the coupling matrix, related to the eigenfrequencies by $\epsilon_k^2= \Lambda_{k}$. From the real symmetry of $E$, the eigenvalues and eigenvectors are real. Furthermore, since we assume the same $\Gamma$ at every node, we obtain for $\Gamma \neq 0$, the same eigenmodes $\vec{c}_k$ with two 
                complex eigenfrequencies $\Omega_{k, \sigma_k=\pm} =  -j \Gamma
              +\sigma_k j \sqrt{\Gamma^2 - \epsilon_k^2 }$. For $\epsilon_k \geq \Gamma$,  $\Im (\Omega_{k})=-\Gamma$. For $\epsilon_k < \Gamma$,  we obtain $\Im (\Omega_{k,-})<-\Gamma$, which produces the fastest amplitude decay and $\Im (\Omega_{k,+})>-\Gamma$, which produces the slowest amplitude decay, creating long-lasting perturbations. 
               Since slowly decaying modes may increase the impact of 
                    disturbances on the  power system stability,
                it  is highly important to 
                find out 
          the topological and system conditions for  such  slow amplitude decays.  \newline

        On the other hand, the stationary state of Eq.(\ref{dynamic}) can also account for the grid topology when written in matrix form,
         
           \begin{eqnarray}\label{steady_nonlin}
           P= KB\sin(B^T\theta).
           \end{eqnarray} 
         
         A DC approximation of the angular differences $\left(\text{i.e.}~ B^T \theta<< \frac{\pi}{2}\right)$ considerably reduces the computational time \cite{overbye} when compared to other more accurate methods such as solving the coupled nonlinear swing equations Eq.(\ref{dynamic}) or solving Eq.(\ref{steady_nonlin}) via a root-finding algorithm. Since $B$, of size $(N,\kappa)$, is related to $L$, $L=BB^T$ (and $L$ to $A$, $L=D-A$), Eq.(\ref{steady_nonlin}) can be expressed as $P=KBB^T\theta$ or $P=KL\theta$ in a DC approximation. For a fully connected graph, $L$ is non-invertible (it possess one zero eigenvalue). Therefore, the steady state phases are obtained from the Moore-Penrose pseudoinverse:

          \begin{equation}\label{steady_lin}
           \theta= \frac{1}{K}L^+P,
           \end{equation} 
where $L^+=(L^TL)^{-1}L^T$. This approximation is accurate enough as long as $P_i<< d_i K$. The reason is that
           $P_{avg_i}=\frac{P_i}{d_i}$  is the average mechanical power generated or consumed at node $i$, which leaves or enters the node in the form of electric power through the transmission lines that are connected to it. If we consider that the electric power in each of these lines does not deviate much from the average, then $P_{avg_i}\approx K\sin(\theta_i-\theta_j$), and since the condition $|\theta_i-\theta_j|<< \frac{\pi}{2}$ is needed for linearization, then $\frac{P_{avg_i}}{K}<<1$ must be fulfilled. The statement $P_i<< d_i K$ follows.

 \section{Electric Power Transmission Grid Models}
        
         Authors in \cite{watts_strogatz} proposed a model that interpolates between a lattice and a random graph based on $p$. For a certain range of $p$, there is a coexistence of small average Path Length, $l_{avg}(p)$, and high average Clustering Coefficient, $C_{avg}(p)$, forming the Small-world network, which mimics many real-world networks that contain small average path lengths, but also have unusually large clustering coefficients \cite{barabasi}. The average Path Length is defined as $l_{avg}(p)= \frac{1}{N(N-1)}\sum_{i,j}l_{ij}$. The Clustering Coefficient is a ratio between the actual number of edges among the neighbors of node $i$ and the number of edges that would exist if those neighbors were fully connected among themselves. Mathematically, $C_i= \frac{h_i}{\frac{1}{2}(d_i)(d_i-1)}$. The average Clustering Coefficient is simply $C_{avg}(p)= \frac{1}{N}\sum_i C_i$. 
         
         A $p=1$ generates a Random network, which may not be necessarily similar to the Erd\"os-R\' enyi random network commonly referred to in the literature. To be more precise, despite similar clustering coefficients and average path lengths, a Watts-Strogatz network with $p=1$ is not identical to an  Erd\"os-R\' enyi random network with same size and same $d_{avg}$, since for example, the Watts-Strogatz algorithm does not allow nodes to exist with degree smaller than $\frac{d_{avg}}{2}$, whereas Erd\"os-R\' enyi does \cite{brenton}.
         
         Here, we study Small-world and Random networks. Firstly, there is a considerable amount of transmission grids that present similar characteristics to the former: Sweden, Finland, Norway, part of Denmark, U.S. Western States, Shanghai, Italy, France, Spain \cite{cuadra} and Northern China \cite{hu}. Secondly, Small-world networks have  economical and structural feasible features for electricity distribution in smart grids, as proven by using real data from the Dutch power grid \cite{pagani}. On the other hand, Random networks are proven to be more robust than multiple networks against intentional attacks \cite{cuadra}, which makes their inclusion also important for our study.


     \begin{figure}[h]
   \includegraphics[totalheight=\valuegiv\textheight, width=\columnwidth]{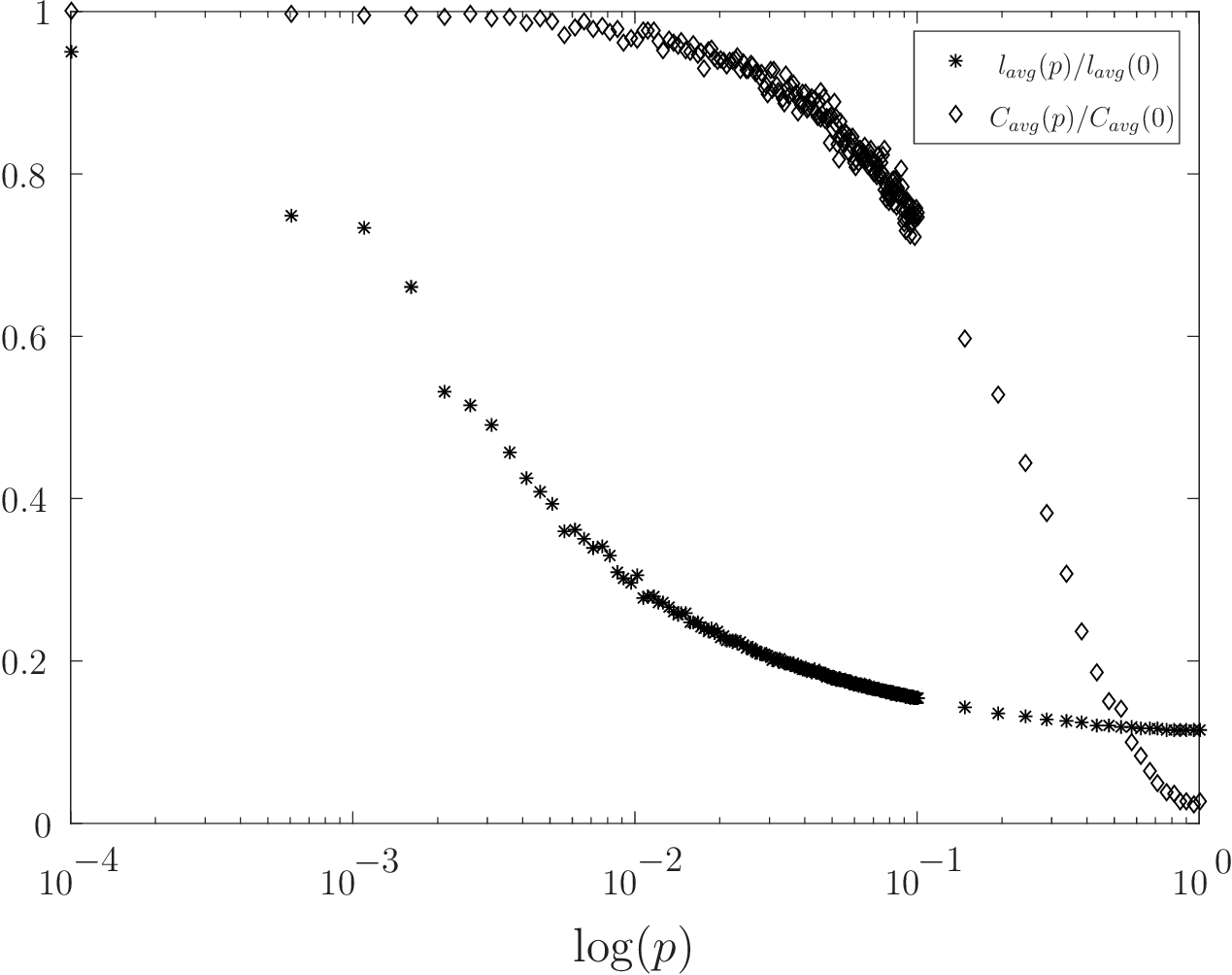}
   \caption{Normalized $l_{avg}(p)$ and $C_{avg} (p)$, with $l_{avg}(0)=25.40$ and $C_{avg}(0)\approx 0.67$,  for the Watts-Strogatz model, with $N=500$ and $d_{avg}=10$. Twenty-five realizations per $p$.}  
   \label{fig:apl_acc}
    \end{figure}
         
\begin{figure*}[htp]
 \centering
 \subfloat[Generators are represented as circles and consumers as crosses. Power line capacity of the network, $K_{sworld}\approx 5.24$ GW. Smallest nonzero eigenfrequency, $\epsilon_2= 5.5470$ Hz.]{\includegraphics[trim = 28.2mm 13.5mm 18mm 22mm, clip, totalheight=\valuegiv\textheight, width=\columnwidth]{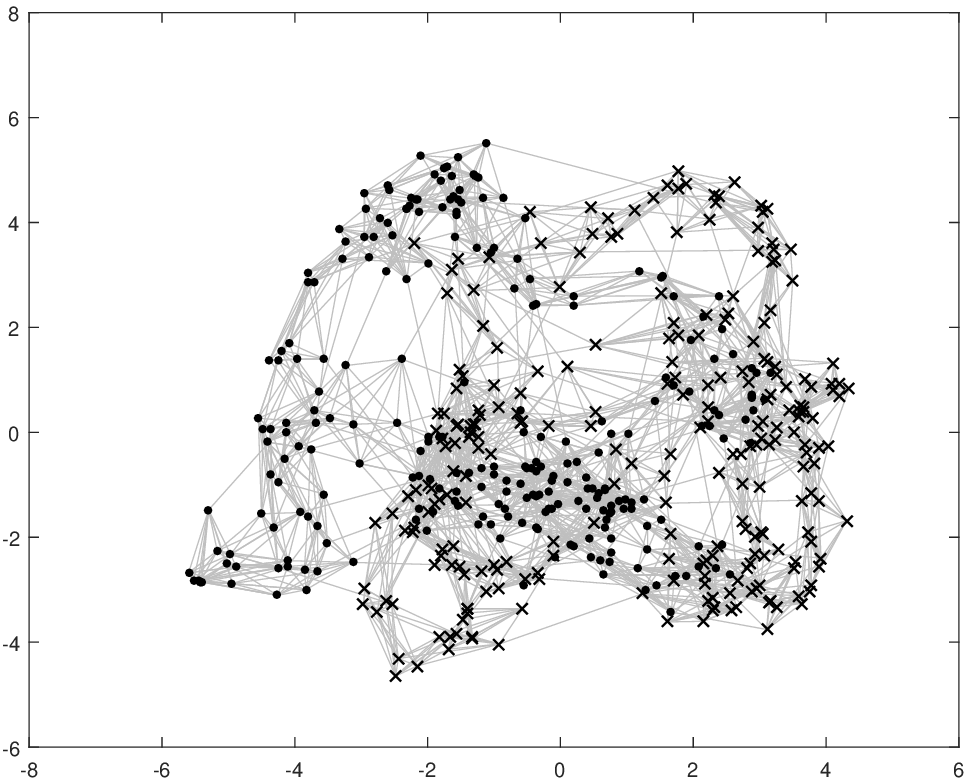}\label{fig:sw}}\hfill
   \subfloat[Randomization of generator and consumer clusters. The squares are those machines that have switched to the opposite power in comparison to Fig. (\ref{fig:sw}). Power line capacity of the network, $K_{sworld} \approx 1.12$ GW. Smallest nonzero eigenfrequency, $\epsilon_2 \approx 2.7198$ Hz.  ]{\includegraphics[trim = 28.2mm 13.5mm 18.2mm 22mm, clip, totalheight=\valuegiv\textheight, width=\columnwidth]{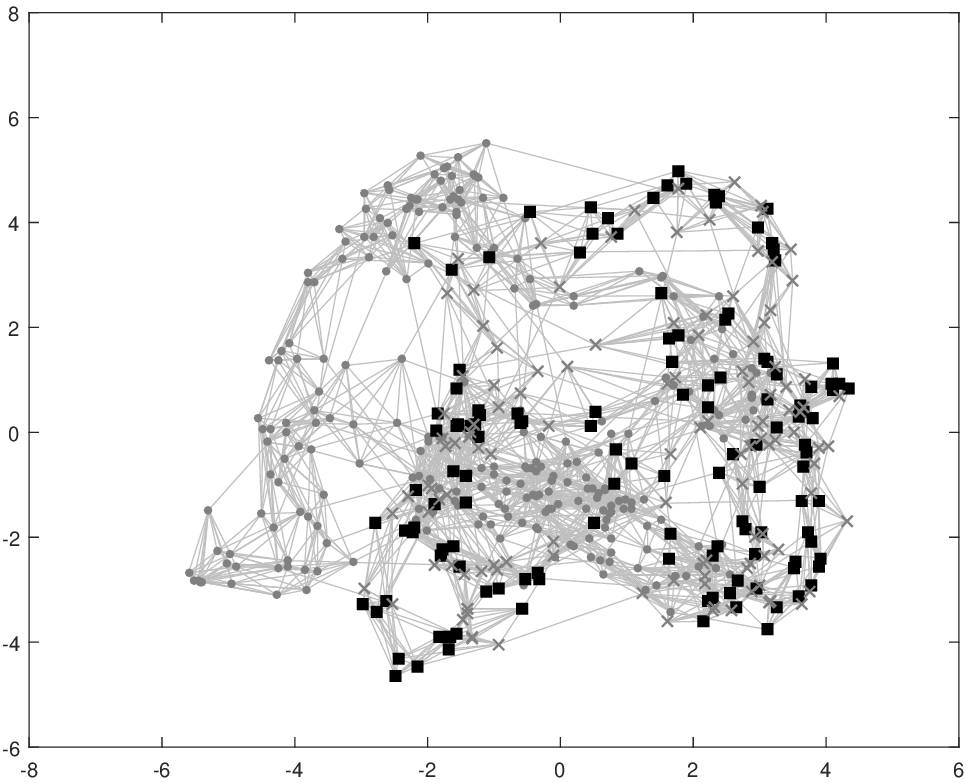}\label{fig:sw1}}\hfill
  \subfloat[Bars size, $0.125$ Hz. Smallest value of $\epsilon_2$ found after $1500$ iterations, $\epsilon_2=2.3764$ Hz.]{\includegraphics[trim = 0mm 0mm 0mm 0mm, clip, totalheight=\valuegiv\textheight, width=\columnwidth]{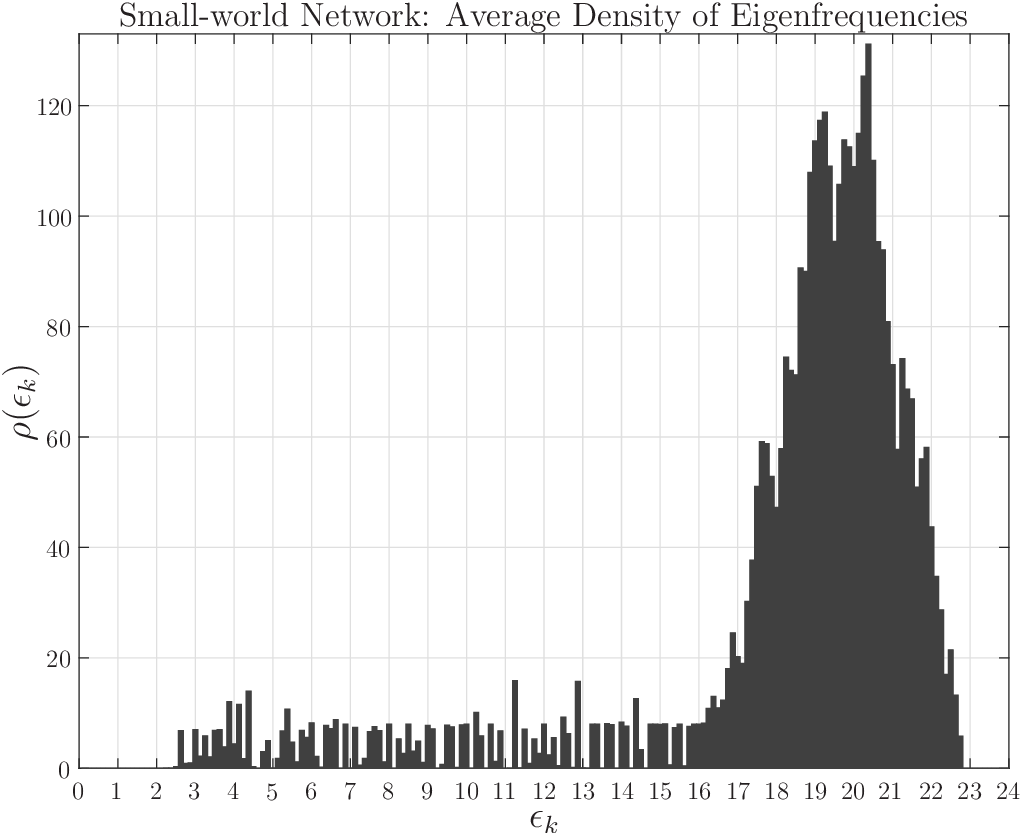}\label{fig:sw2}} \hfill
   \subfloat[Intensity of the generalized Fiedler vector components (i.e $|c_{i2}|^2/\max_i|c_{i2}|^2$) for the grid in Fig. (\ref{fig:sw1}).]{\includegraphics[trim = 10.5mm 13mm 18.2mm 19mm, clip, totalheight=\valuegiv\textheight, width=\columnwidth, clip, totalheight=0.30\textheight, width=\columnwidth]{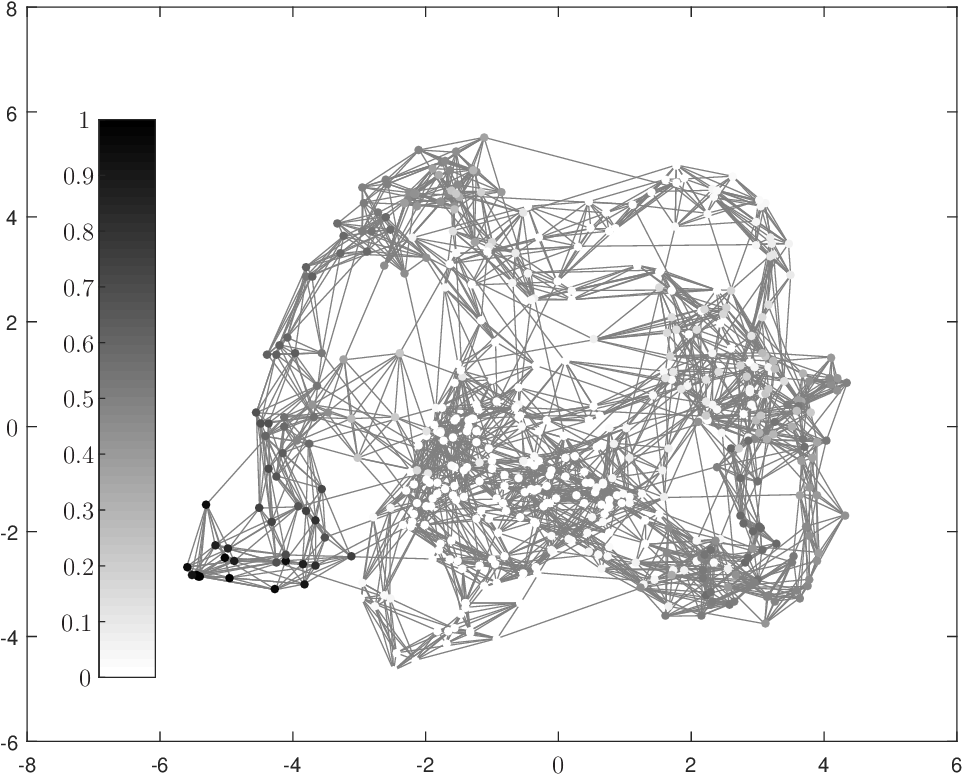}\label{fig:sw3}} \hfill
  \caption{Small-world Network from the Watts-Strogatz Model with Parameters: $N=500$, $d_{avg}=10$, $p=3.42*10^{-2}$, $l_{avg}(p)/l_{avg}(0)=0.198$, $C_{avg}(p)/C_{avg}(0)=0.923$, $l_{avg}(0)=25.40$ and $C(0)\approx 0.67$.}
\end{figure*}

\begin{figure*}[!t]
 \centering
 \subfloat[Generators are represented as circles and consumers as crosses. Power line capacity of the network, $K_{rand}\approx 335.14$ MW. Smallest nonzero eigenfrequency, $\epsilon_2= 5.8986$ Hz.]{\includegraphics[trim = 35mm 15mm 15mm 16mm, clip, totalheight=\valuegiv\textheight, width=\columnwidth]{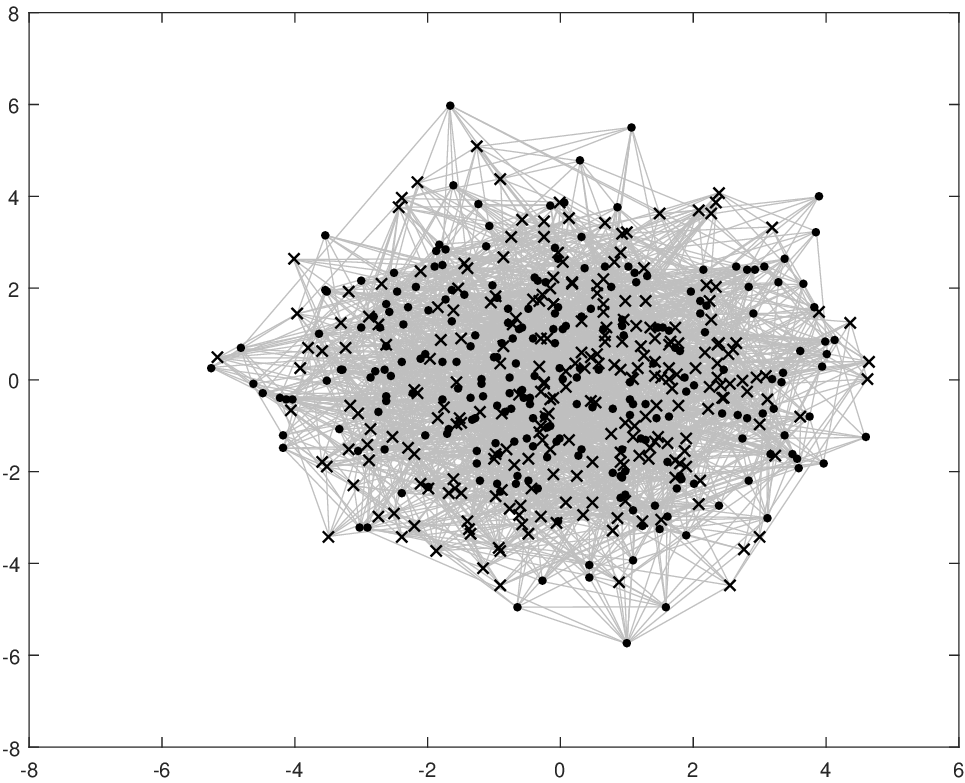}\label{fig:rand}}\hfill
   \subfloat[Randomization of generator and consumer clusters. The squares are those machines that have switched to the opposite power in comparison to Fig. (\ref{fig:rand}). Power line capacity of the network, $K_{rand} \approx 309.34$ MW. Smallest nonzero eigenfrequency, $\epsilon_2 \approx 5.6250$ Hz.]{\includegraphics[trim = 35mm 15mm 15mm 16mm, clip, totalheight=\valuegiv\textheight, width=\columnwidth]{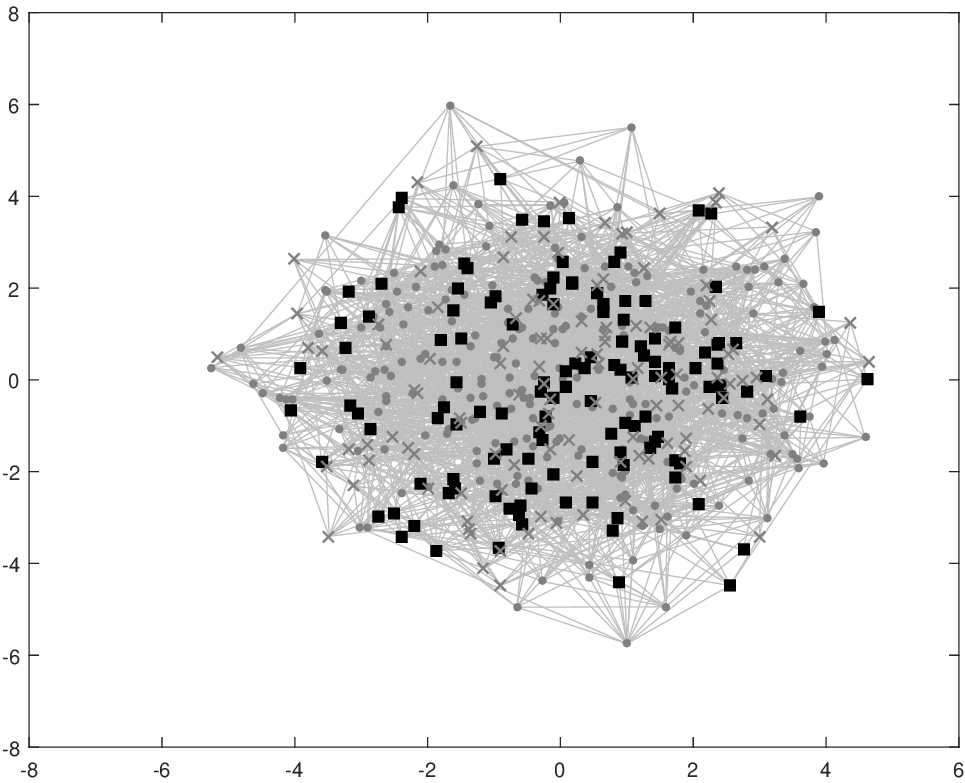}\label{fig:rand1}}\hfill
     \subfloat[Bars size, $0.125$ Hz. Smallest value of $\epsilon_2$ found after $1500$ iterations, $\epsilon_2=5.4880$ Hz.]{\includegraphics[trim = 0mm 0mm 0mm 0mm, clip,totalheight=\valuegiv\textheight, width=\columnwidth]{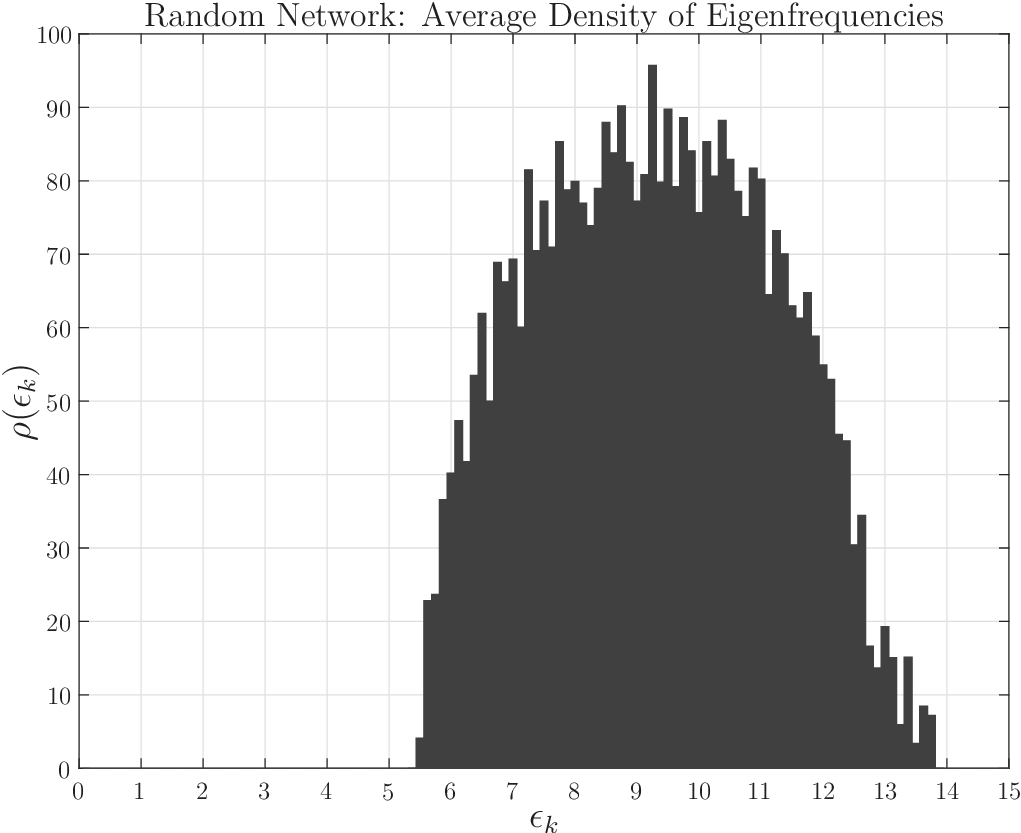}\label{fig:rand2}} \hfill
   \subfloat[Intensity of the generalized Fiedler vector components (i.e $|c_{i2}|^2/\max_i|c_{i2}|^2$) for the grid in Fig. (\ref{fig:rand1}).]{\includegraphics[trim = 13mm 13mm 15mm 12mm, clip, totalheight=\valuegiv\textheight, width=\columnwidth]{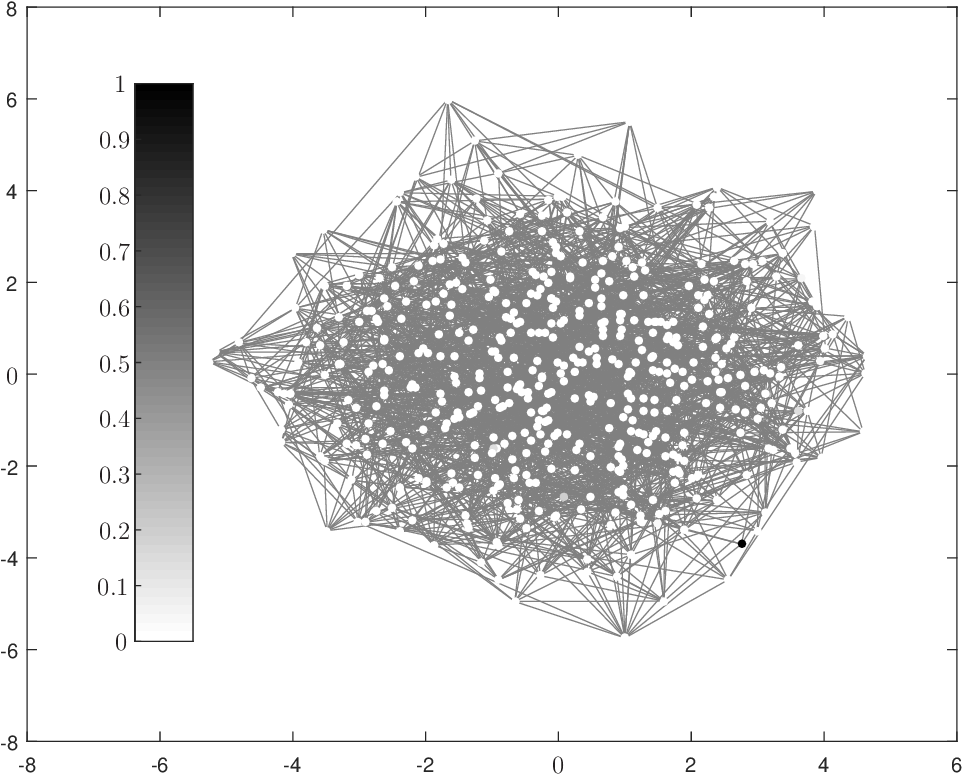}\label{fig:rand3}} \hfill
  \caption{Random Network from the Watts-Strogatz Model with Parameters: $N=500$, $d_{avg}=10$, $p=1$, $l_{avg}(p)/l_{avg}(0)=0.1154$, $C_{avg}(p)/C_{avg}(0)=0.0239$, $l_{avg}(0)=25.40$ and $C_{avg}(0)\approx 0.67$.}
\end{figure*}

         To create our grids, we select $N=500$ and $d_{avg}=10$. For the Small-world network we fulfill the condition $N>> d_{avg} >> \ln (N) >> 1$, to have a sparse but connected graph \cite{barabasi}. Then, we find $l_{avg}(p)$ and $C_{avg}(p)$, for different values of $p$. For each $p$, we average $l_{avg}(p)$ and $C_{avg}(p)$ over 25 realizations. We compare both normalized parameters in Fig. (\ref{fig:apl_acc}), and then retrieve the $p$ that yields the greatest difference between them. This rewiring probability is $p=3.42*10^{-2}$, with normalized parameters $l_{avg}(p)/l_{avg}(0)=0.198$ and $C_{avg}(p)/C_{avg}(0)=0.923$, which results in the Small-world network shown in Fig. (\ref{fig:sw}).  For the Random network, we set $p=1$, which results in the grid shown in Fig. (\ref{fig:rand}), with $l_{avg}(p)/l_{avg}(0)=0.1154$ and $C_{avg}(p)/C_{avg}(0)=0.0239$. The parameters for $p=0$ are $l_{avg}(0)=25.40$ and $C_{avg}(0)\approx 0.67$.


\begin{figure*}[!t]
 \centering
 \subfloat[Generators are represented as circles and consumers as crosses. Power line capacity of the network, $K_{G}= 10$ GW.]{\includegraphics[trim = 25mm 14mm 25mm 12mm, clip, totalheight=0.36\textheight, width=\columnwidth]{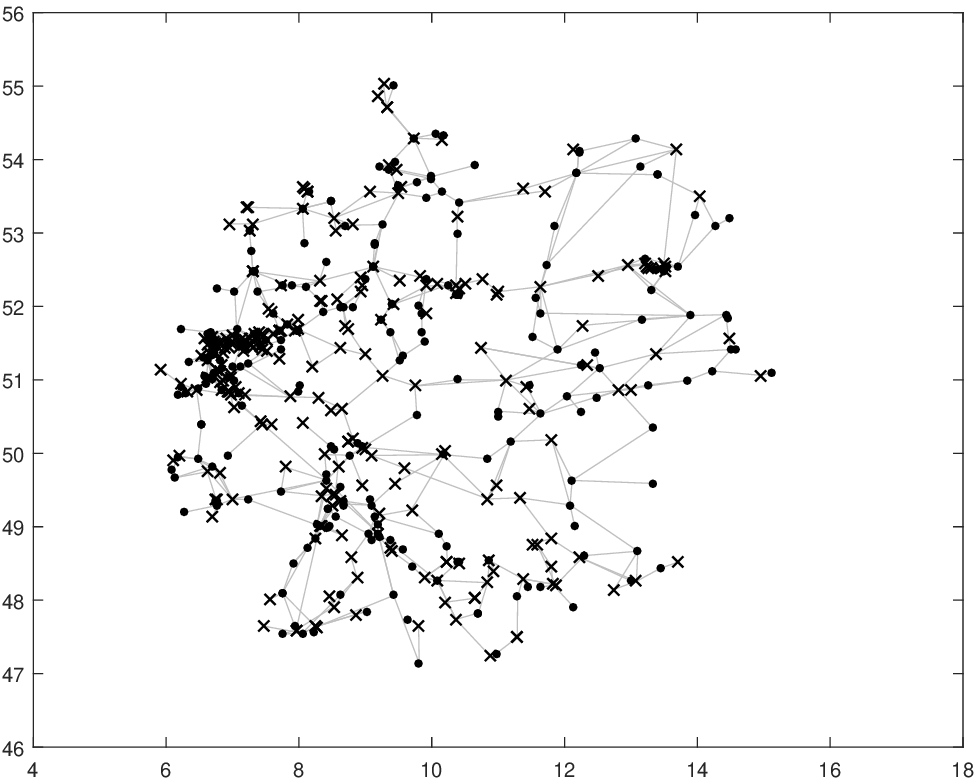}\label{fig:ger}}\hfill
 \subfloat[Intensity of the  generalized Fiedler vector components (i.e $|c_{i2}|^2/\max_i|c_{i2}|^2$) for the grid in Fig.(\ref{fig:ger}).]{\includegraphics[trim =25mm 14mm 12mm 12mm, clip, totalheight=0.36\textheight, width=\columnwidth]{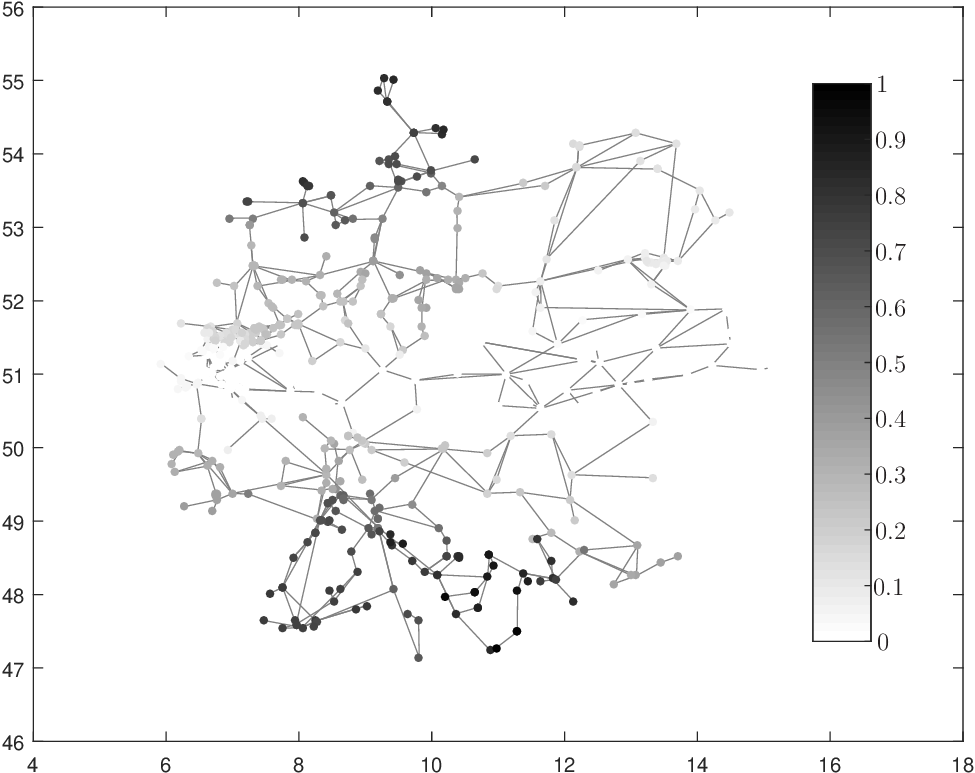}\label{fig:ger3}}\hfill
     \subfloat[Bars size, $0.5$ Hz. Smallest value of $\epsilon_2$ found after $1500$ iterations, $\epsilon_2=1.7343$ Hz. Peak density around $\rho_{peak}= \sqrt{\frac{K_G}{J\omega}}= 17.84$ Hz.]{\includegraphics[trim = 0mm 0mm 0mm 0mm, clip, totalheight=\valuegiv\textheight, width=\columnwidth]{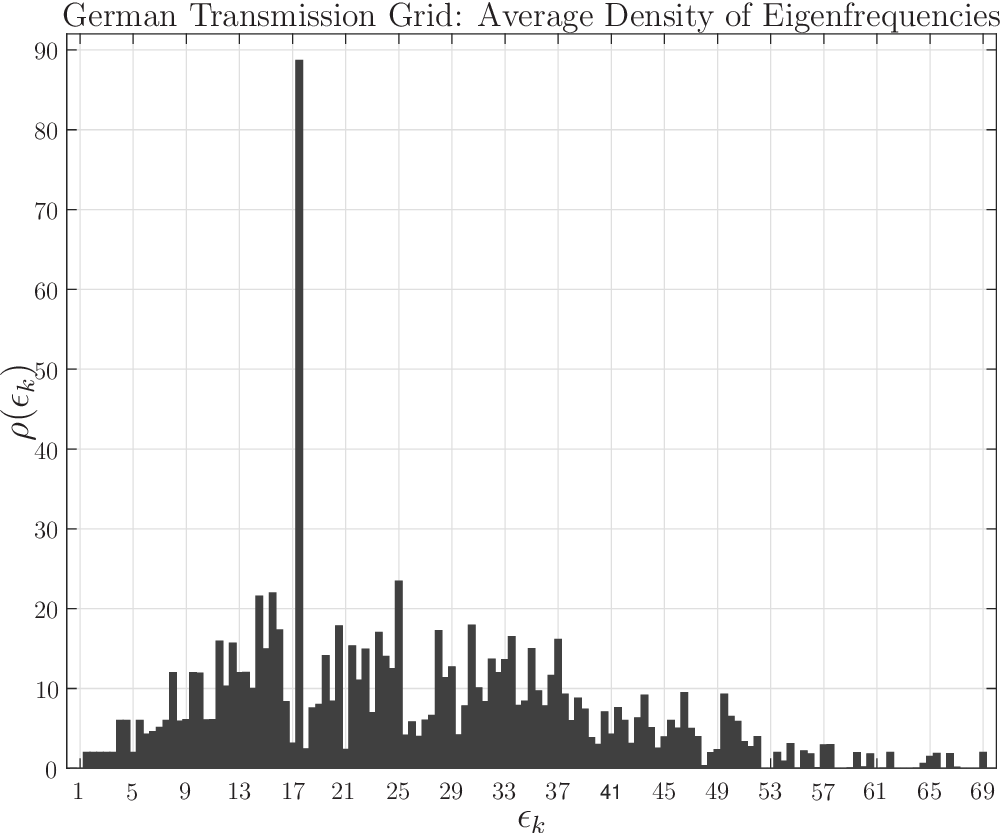}\label{fig:ger2}} \hfill
  \caption{Extra-high-AC Voltage (380 kV and 220 kV) German Transmission Grid with Parameters: $N=489$, $d_{avg}=2.71$, $l_{avg}=9.9384$, $C_{avg}=0.2021$.}
\end{figure*}

        We also consider the fully connected graph of the Extra-high-AC Voltage (380 kV and 220 kV) German transmission grid, which can be found in \cite{scigrid}. The grid, consisting of 489 nodes, is shown in Fig. (\ref{fig:ger}), with parameters $l_{avg}=9.9384$ and $C_{avg}=0.2021$.
       
  \section{Spectral Analysis}
   
          \subsection{Generalized Laplacian Matrix} 
          

      Authors in  \cite{dorfler} showed that a sufficient condition for Small-world and Random networks to reach cohesive phases (that is, that all angular distances $|\theta_i- \theta_j|$ are bounded $|\theta_i- \theta_j| \leq \zeta < \frac{\pi}{2}$,
       where the upper bound is known from the power-angle curve of a synchronous generator connected to an infinite busbar), is given by $||B^{T}L^{+}P||_{\infty}\leq K\sin(\zeta)$. In the limit $\zeta -> \frac{\pi}{2}$, we find:

 \begin{equation} \label{syn_cond}
  ||B^{T}L^{+}P||_{\infty}< K.
 \end{equation}
 
 If we compare Eq.(\ref{steady_lin}) and Eq.(\ref{syn_cond}), we observe that $||B^{T}\theta||_{\infty}< 1$. This implies that $\max_{i,j}|\theta_i-\theta_j|<1$ (approx. $57.30^o$). If Eq.(\ref{syn_cond}) is imposed, then $E$ is a diagonally dominant matrix. A diagonally dominant matrix $W$ must satisfy $|W_{ii}|\geq\Sigma_{j\neq i} |W_{ij}|$. Since $\max_{i,j}|\theta_i-\theta_j|<1$, the equality $|E_{ii}|=|- \Sigma_{j \neq i} E_{ij}|=\Sigma_{j \neq i} |E_{ij}|$ holds. This makes $E$ a diagonally dominant matrix with positive diagonal entries. Therefore, $E$ is a positive semi-definite matrix, just as $L$. 

In conclusion, the coupling matrix $E$ is a positive semi-definite and real-symmetric matrix; thus, its eigenvalues are real and non-negative, indicating that the system always reaches steady-state under phase-cohesiveness. Moreover, $E$ has positive entries along the diagonal, negative entries for adjacent nodes and zeros for nonadjacent nodes; therefore, it can be considered as a generalized Laplacian matrix. These matrices, which include the graph Laplacian and are found in the inverse eigenvalue problem of a graph \cite{hogben}, precisely fulfill the same conditions for the off-diagonal entries, but have no restrictions on the diagonal entries. Beyond that, the coupling matrix can be considered as nothing else than the Laplacian matrix of a positively weighted graph. Besides, all properties of $L$ apply to $E$, including the decomposition into a product of oriented incidence matrices, i.e $E= QQ^T$, where $Q_{iu}=\sqrt{\frac{K}{J\omega}\cos(\theta_i-\theta_j)}$ ($i$ as source node) and $Q_{ju}=-\sqrt{\frac{K}{J\omega}\cos(\theta_i-\theta_j)}$ ($j$ as sink node) and where $u$ is in the edge set.
 

 \subsection{Algebraic Connectivity}
 
 The smallest nonzero eigenvalue of the Laplacian matrix is called the algebraic connectivity, and its corresponding eigenvector, the Fiedler vector \cite{fiedler}. Since the coupling matrix depends on the angular differences, which are inherently related to the power distribution, it would be highly convenient to know lower and upper bounds 
 for its respective generalized  algebraic connectivity $a_E(G)$ to avoid performing an eigenvalue decomposition every time $P$ changes. The $a_L(G)$ of the matrix $\frac{K}{J\omega} L$ can be set as the upper bound for $a_E(G)$. The smaller the angular differences, the more the coupling matrix approaches the scaled Laplacian $\left(\text{i.e.}~ E \rightarrow \frac{K}{J\omega}L \right)$, and the closer $a_E(G)$ gets to $a_L(G)$. The matrix $\frac{K}{J\omega} L$ corresponds to the coupling matrix of a network with $P=0$, where the only power in the grid is that of the perturbation itself. It would also be very useful to provide
  lower bounds for the generalized algebraic connectivity of $E$. Some lower bounds have been derived for weighted graph Laplacian matrices \cite{kirkland, bermann}, but the inclusion of, for instance, the weighted isoperimetric number \cite{bermann}, makes their calculation computational expensive  in comparison to 
   the explicit solution of the  eigenvalue problem, which  we perform in the following section. 
 




\section{Simulation}

    
\subsection{Selection of Grid Parameters}
   
   So far we have described multiple ways in which system variables can be directly related to topology. A strong emphasis has been given to $E$ as it not only contains information about the topology but also about the system operating state. In real power systems, those variables (node power, power line capacity, inertia, etc.) widely change depending on, for instance, the location of power generating sources and loads. These can exert different effects on mode distributions when small perturbations occur; thus, results from one system may not precisely apply to another one. Therefore, as previously mentioned, we prioritize parameter homogeneity to capture the influence that topology may have when the system interacting agents have opposite behaviors (i.e. producers and consumers).  
 
  To be more specific, we assign values of power to each node from a bipolar distribution, i.e $P_i=\pm P$ in Watts. $P>0$ for generators and $P<0$ for motors (consumers). Eq.(\ref{dynamic}) synchronizes at a frequency $\dot{\theta}_{\text{synch}}= \sum_i^N \frac{P_i}{\gamma}$ \cite{dorfler}, which implies that the condition $\sum_i^N P_i=0$ must be fulfilled at all times for the system to reach steady state. This is a realistic consideration since power generation must constantly match load demand. We take as reference the German installed capacity of $199.2$ GW as per November $10^{th}$, $2015$ \cite{ministry}, and consider half of the nodes-for an even number- to be generators and the remaining half to be consumers. For the $500$-node complex networks (Small-world and Random) this results in $P_i= \pm 796.80$ MW. For the $489$-node German grid, we have on average $P_i\approx \pm 814.72$ MW. We choose the grid angular frequency $\omega= 2\pi$($50$ Hz) and moment of inertia $J=10^5$ kgm$^2$. This $J$ is, for instance, for a generator working  at $\omega$, with inertia constant $H=3 \frac{MJ}{MVA}$, rated at $100$ MVA. This $H$ resembles those of high-speed and slow-speed water-wheel generators and non-condensing turbine generators \cite{birron}. We apply Eq.(\ref{syn_cond}) to the complex networks and retrieve $K_{sworld}\approx 5.24$ GW and $K_{rand} \approx  335.14$ MW. For the German transmission grid, we assign $K_{G}=10$ GW, which is strong enough to keep small angular differences. Finally, we select the damping rate $\Gamma=1$ Hz for all topologies considered.
    %
    \subsection{Density of Eigenfrequencies} 
       
        We study the eigenfrequency density $\rho_m(\epsilon)$ for different arrangements of generators and consumers by randomizing, within the bipolar distribution, the $P_i$ of each node. We perform $R=1500$ iterations to obtain the average density, $\rho(\epsilon)= \frac{1}{R}\sum_{m=1}^{R}\rho_m(\epsilon)$. The results are shown in Figs. (\ref{fig:sw2},\ref{fig:rand2},\ref{fig:ger2}). The stationary solution corresponds to $\epsilon_1= 0$ Hz and it is not shown. 
        
        Due to the parameter homogeneity (precisely, $J$ and $K$) and the very small angular differences, it is easily perceivable that the eigenfrequencies plots are very close to the scaled versions- by a factor of $\frac{K}{J\omega}$ - of their corresponding $L$-spectra. This means that they represent the influence of topology alone when the power vector is binarily distributed throughout the grid. In such circumstances, we observe that:
  
    \begin{itemize}
    \itemsep0em 
    \item The nonzero eigenfrequencies for all networks exceed, for the chosen parameters, the damping rate $\Gamma$ (i.e $\epsilon> \Gamma$), so that disturbances decay exponentially fast with relaxation rate $\Gamma$. We currently know that under high integration of renewable energy, the system inertia will be considerably reduced \cite{ulbig}. With homogeneous parameters, a decrease of $J$ increases $\Gamma$ much more than $\epsilon$ since $\Gamma \propto \frac{1}{J}$ whereas $\epsilon \propto \frac{1}{\sqrt{J}}$. A considerable decrease of inertia shall produce at least one long-lasting perturbation mode (i.e. $\Gamma>\epsilon_2$) in the system.
    \item Although highly distributed, a significant peak of the German grid eigenfrequency density is located around $\sqrt{\frac{K_{G}}{J\omega}} \approx 17.84$ Hz. This could be attributed to the influence of the 149 one-degree nodes in the grid whose diagonal entries in $E$ are, with very small angular differences in the system, $\sqrt{\frac{K_{G}}{J\omega} \cos(\theta_i-\theta_j)} \approx \sqrt{\frac{K_{G}}{J\omega}}$.
    \item The average eigenfrequency density of the Random network resembles the Mar\u{c}henko-Pastur distribution, expected for uncorrelated random matrices. 
    \end{itemize}
    
    \subsection{The Effect of Clustering}
    
  The values $K_{sworld}$ and $K_{rand}$ were obtained for the vectors $P_{sworld}$ and $P_{rand}$ assigned to Figs. (\ref{fig:sw},\ref{fig:rand}), in which there are visible clusters of generators and consumers. If $P$ is randomized, reducing the size of the clusters, smaller values of $K$ can be found. This effect was studied in \cite{buzna} for a bipolar distribution of frequencies (power in our case), and it was shown that synchronization is enhanced when adjacent nodes have opposite frequencies, resulting in a diminished frequency similarity throughout the grid. This simply means that synchronization is enhanced when generators are surrounded by consumers and vice versa. Moreover, critical effects, such as cascading failures, are less likely to be triggered if a greater frequency dissimilarity prevails. This was demonstrated statistically in \cite {jung}, where authors claim that the existence of large clusters of generators and consumers turns the grid vulnerable against cascading failures, since the likelihood for a whole cluster to disconnect at once appears to increase with increasing cluster size. 
 
 Figs. (\ref{fig:sw1},\ref{fig:rand1}) provide an insight into the effect of randomization. The squares are those machines that have switched their power in comparison to Figs. (\ref{fig:sw},\ref{fig:rand}). It is clear that clusters are reduced, resulting in smaller power line capacities (i.e $K_{sworld} \approx 1.12$ GW and $K_{rand} \approx 309.34$ MW), but also in  smaller nonzero eigenfrequencies. For the Small-world network, we obtained for Fig. (\ref{fig:sw}), $\epsilon_2=5.5470$ Hz, whereas for Fig. (\ref{fig:sw1}), $\epsilon_2=2.7198$ Hz. For the Random network, we obtained for Fig. (\ref{fig:rand}), $\epsilon_2= 5.8986$ Hz, whereas for Fig. (\ref{fig:rand1}), $\epsilon_2= 5.6250$ Hz. In fact, out of the $1500$ iterations for each complex network, no single value of $\epsilon_2$ was greater than the ones from Figs. (\ref{fig:sw},\ref{fig:rand}). 
 This implies that whereas clustering is detrimental to grid stability and to cascading outage prevention, the larger power capacity needed to ensure stability in the presence of clusters results in an increment of the smallest nonzero eigenfrequency, leading in fact to faster mode damping rates and thereby to a greater resilience of the power system to small perturbations.
 
  \subsection{Spatial Distribution of the Generalized Fiedler Vector Intensity}
   In Figs. (\ref{fig:sw3},\ref{fig:rand3},\ref{fig:ger3}), we show the intensity, $|c_{i2}|^2$, of the generalized Fiedler vector; the eigenmode corresponding to the smallest nonzero eigenfrequency. The intensity at each node is 
    divided by $\max_i |c_{i2}|^2$.
    We observe that in the Random network, Fig. (\ref{fig:rand3}),
     the eigenmode is strongly localized with most of its intensity on a single node. 
     In the Small-world network,  Fig.  (\ref{fig:sw3}), the eigenmode intensity is 
   spread over many nodes, which are far away from each other. In the German transmission grid, 
       the intensity is spread over most of the grid, with greater intensity in the Southern and Northwestern part of the system, see Fig. (\ref{fig:ger3}). 
       To understand this behavior, we can refer to the fact that the discrete wave equation 
         Eq.(\ref{alphaharmonic4}) was first derived 
         for the problem of randomly coupled atoms
                 in  harmonic approximation and has been intensively studied   
                  for various random distributions of the coupling $t_{ij}$ \cite{dyson,alexander,ziman,wegner}.            
        For nonzero eigenfrequency  $\epsilon_k$, 
               the eigenmodes were found,  for a random chain of nodes,
                to be localized with  localization length  
                $ \xi(\epsilon_k) \sim 1/\epsilon_k$ \cite{dyson,alexander,ziman,wegner,john},
                 due to the random scattering of waves along the chain. This is an example of the 
                  so-called
                  Anderson localization, which  
                  is enhanced when the amplitude of  randomness is increased \cite{anderson}.
                 In grids with higher $d_{avg}$, the localization length is typically found to be larger.                 
                     Moreover, 
                      the localization length is typically smallest in tree-like grids, whereas it becomes 
                       larger the more meshed the grid is; in which case the eigenmodes 
                         can become even delocalized  \cite{wegner, john}. On the other hand, $C_{avg}$ is a measure of 
                         how strongly meshed a grid is. 
                     We observe that  the Random network, Fig. (\ref{fig:rand3}),
                       has a  very small average clustering coefficient $C_{avg}=0.016$, 
                       explaining the fact that its eigenmode is strongly localized, 
                        whereas the  German transmission grid, shown in Fig. (\ref{fig:ger3}), is  meshed
                      with  $C_{avg}= 0.2021$  and  the Small-world network in 
                       Fig.   (\ref{fig:sw3}) is more strongly meshed  with $C_{avg}= 0.61841$, explaining that the eigenmode intensity 
                        in these grids is more delocalized and spread over  many nodes.     
The observation 
      that the eigenmode is strongly localized in the Random network 
      may have important consequences for the design of stable electricity grids: 
                       if  the phase perturbation  is initially in  a  state localized 
                       around a node 
                        with localization length $\xi_k$, then that disturbance  remains 
                         localized there and it 
                        decays exponentially  in time \cite{kettemann}. Thus, lesser meshed grids  may help to localize 
           disturbances more strongly.

 
 



\section{Potential Contributions to Grid Control Strategies}

{\it Dynamic Transmission Topology Control (TC)}: Although scalable practical solutions have not been yet achieved, there is an ongoing interest in the research community for this emerging control technology \cite{sari} for its potential to manage the uncertainty of power sources and flow patterns on a grid with high penetration of renewable energy \cite{wang}. The control strategy consists of switching lines on or off to relieve voltage and line flow violations \cite{wang}, with a considerable impact on small-signal and transient stability \cite{sari, wang}. Based on our analysis, any TC action that modifies the structure of the grid should firstly guarantee phase-cohesiveness (e.g. through grid mechanical power tunning \cite{fazlyab} as an optimization problem) and secondly that $\sqrt{a(L')}$, where $L'$ is the Laplacian matrix of the modified topology, is never less than the damping rate of the system; otherwise, there will surely exist at least one mode that decays slowly in time. \newline

{\it Power System Stabilizer:} Here, we highlight the benefit of providing a mode-related input to a PSS, which is the main grid control device to guard the system against small-signal instability. PSS are off-line tuned generator controllers for which significant disadvantages have been found, mainly due to being local devices that do not use remote signal inputs and therefore do not adaptively change their setpoints according to the power system operating conditions \cite{derong}. 

Formerly, PSS used a measurement of the speed deviation of a number of points along the generator's shaft to then calculate the average speed deviation. For long shafts prone to torsional oscillations, this method turned out to be troublesome \cite{machowski}. It was later found that the need to measure the speed deviation at a number of points along the shaft can be avoided by calculating the average speed deviation from measured electrical quantities. This method indirectly calculates
the equivalent speed deviation $\Delta\omega_{i}^{eq}$ from the integral of the accelerating power at machine $i$ \cite{machowski}:

 \begin{equation} \label{pss1}
       \Delta\omega_{i}^{eq}(t)= \frac{1}{J \omega} \int_{-\infty}^{t} {\left(\Delta P_{i}(t')- \Delta P_{e_i}(t')  \right)dt'},
    \end{equation}
where $\Delta P_{i}$ and $\Delta P_{e_i}$ are power changes at node $i$. The former can be  retrieved from the angular frequency measured by the end-of-schaft speed sensing system \cite{machowski}.
  Thus, this PSS requires  two local input signals. Nonetheless, we subsequently show that $\Delta\omega_{i}^{eq}(t)$ can account for the topology and system state  once the eigenmodes of the coupling matrix are known. From Eq.(\ref{alphaharmonic3}),
  we note that  $( \Delta P_{i}(t)- \Delta P_{e_i}(t))/(J \omega) =\partial_t^2   \alpha_i  + 
                  2 \Gamma  \partial_t   \alpha_i.
  $ 
  Therefore, we find that  $ \Delta\omega_{i}^{eq}(t)= \partial_t  \alpha_i (t)  + 
                  2 \Gamma   \alpha_i (t)$. 
                  Next, we can insert the expansion of  
                  the phase deviations 
                   in terms of the eigenvectors  $\vec c_k$ of the coupling matrix,
                   $\alpha_i (t)   =  \sum_{k=1, \sigma = \pm}^N b_{k \sigma}
               c_{i k} \exp {(-j \Omega_{k \sigma} t)}$. Thereby, we find:

\begin{equation} \label{pss2}   
	\Delta\omega_{i}^{eq}(t)= 
	 \sum_{k, \sigma_k} (- j \Omega_{k \sigma_k} +2 \Gamma)c_{ik} b_{k \sigma_k}\exp{(-j\Omega_{k \sigma_k} t)}.  
\end{equation}

    Then, it remains to find the expansion coefficients $b_{k \sigma_k}$ in response to 
     changes of electric power. 
      This has been recently obtained as a spectral 
       representation of the linear response to 
     changes in $P_{e_i}$ \cite{tamrakar} 
      and in terms of a weighted  integral over time of 
      the change in $P_{e_i}$ \cite{haehne}.
      Employing these results, we finally get: 
      
      \begin{eqnarray} \label{pss3}   
	&&\Delta\omega_{i}^{eq}(t)= 
	 \sum_{k, \sigma_k} \frac{\sigma_k (- j \Omega_{k \sigma_k} +2 \Gamma) c_{ik} }{\sqrt{ 1 - \tau^2 \epsilon_k^2}}  \times\nonumber \\   
	&& \int_{-\infty}^t \frac{d t'}{\tau} \exp{(-j\Omega_{k \sigma_k} (t-t'))} \sum_v c_{vk}\Delta P_{e_v} ( t'),
\end{eqnarray}
where $\tau = 1/\Gamma$. As noted, only $\Delta P_{e_i}(t')$ is now needed as an input signal. Furthermore, knowing the eigenvectors $\vec c_k$ allows 
    to reduce the number of nodes $v$ from which 
    the input signal $\Delta P_{e_v} ( t')$ is needed. This is because, since the generalized Fiedler vector has the largest impact on grid stability, the PSS may only need to take  signals from those nodes with greatest eigenmode intensity. A proper design for a PSS to implement Eq.(\ref{pss3}) in order to produce a leading voltage signal to control perturbation modes would then be a topic of further discussion.


 \section{Conclusion}

This paper, far from delving into the already thorough studies
of small-signal stability, aims to identify the causes that make perturbations propagate in some way or another. For this purpose, we perform a similar eigenanalysis but establish a direct relationship between the dynamics and
the network structure. We first focus on eigenfrequency
plots to show that the German grid or other real networks
(e.g Italy, France, etc.) with Small-world traits may produce
long-tailed distributions. Of course, in real power grids, system
parameters exert a strong influence on the distributions, but
in order to isolate the influence of topology, we have only
assigned binary powers and homogeneous parameters. Moreover,
with these plots, we have shown that although clusters
are detrimental to the control of critical effects (i.e cascading
failures, synchronization), they make small perturbations fall
faster due to increased power line capacities and therefore
increased damping rates. 

In power system planning, the addition of one node or one edge shall take into account the network's average clustering coefficient in order to improve system controllability, as we have found strong indications that the degree of localization of the generalized Fiedler eigenmode intensity tends to
increase with a decrease of $C_{avg}$. It is, at the same time, crucial to implement strategies to keep track of the generalized algebraic connectivity to avoid having long-lasting perturbations in the system. 

Finally, we have proposed ways in which future and existing grid control strategies can consider the influence of topology in their designs and actions. Therefore, we expect
that, having addressed multiple topological aspects in relation
to perturbations dynamics, this study serves as an insightful
source for upcoming research on power system planning and
control.

\section{Acknowledgment}
  We gratefully acknowledge the support of BMBF
   CoNDyNet FK. 03SF0472A.

\end{document}